\documentstyle[12pt,twoside,fleqn,epsfig]{article}
\newcommand{\be}{\begin{eqnarray}}
\newcommand{\ee}{\end{eqnarray}}

\begin{document}
 
\begin{tabbing}
\`SUNY-NTG-97-17\\
\
\end{tabbing}
\setlength{\parskip}{20pt}
\vbox to  0.8in{}

\centerline{\Large\bf  Event-by-event analysis of heavy ion collisions  }
\centerline{\Large\bf and     thermodynamical   fluctuations  }
\vskip 2.5cm
\centerline{\large
  E.V. Shuryak \footnote{Email: shuryak@dau.physics.sunysb.edu}}
\vskip .3cm

\centerline{Department of Physics}
\centerline{State University of New York at Stony Brook}
\centerline{Stony Brook, New York 11794}
\vskip 0.35in
\centerline{\bf Abstract}
 The
event-by-event analysis of heavy ions collisions is becoming possible
with advent of
large acceptance detectors: it can provide dynamical
information which cannot be obtained from inclusive spectra.
We identify  some observables which can be
related to standard thermodynamical theory of fluctuations
 and therefore may provide information about properties
  of hadronic matter at the freeze-out stage. Among those are fluctuations of
``apparent temperature'' (the $p_t$
slope),  as well as the  population  of different
bins  of the pion momentum distribution.
\indent
\eject     
  1.In the field of  heavy ion collisions the
   event-by-event search for ``unusual'' events (deviating in some way from
   the
average behavior) is attracting increasing attention. Some
future experiments at RHIC (especially STAR and PHOBOS) consider it
to be among the central issues of their physics program.
  
  In general,  deviations from the average behavior
  can be of $dynamical$ or of
$statistical$ nature. A well known example of the former kind
is strong dependence of all observables on
the  value of the impact parameter b\footnote{  Because of its paramount
significance most
experiments are able to fix its value
 already at the trigger level,  by measuring  charge multiplicity, 
  transverse or ``spectator" energy.}.
Furthermore, $direction$ of the impact
parameter $\vec b$ in the $\phi$ angle is important for
  studies of the specific collective effect known as ``directed flow''. 
 Non-trivial deviations of dynamical nature due to QCD phase transition
 were for example predicted in
the AGS/SPS energy domain, due to  formation of
QGP bubbles  \cite{KC} or of the 
disoriented chiral condensate  (DCC) \cite{DCC}. 
In order to see whether those are present or absent in the data, one
should have a proper benchmark, the statistical
fluctuations we discuss below.
 
  Fluctuations  can also be sensitive to non-trivial details of the
dynamics.  An interesting example
  was considered by Gazdzicki and Mrowczynski  \cite{GM_92}:
  event-by-event fluctuations of the  $<p_t>_{event}$ turns out to be sensitive
 to  models describing nuclear collisions. 
In particular, models considering just
 a superposition of independent
pp events (like e.g. the original VENUS event generator without
re-scattering) predict much larger fluctuations compared to 
  models  with significant re-scattering of secondaries (e.g. RQMD). 
 This paper has triggered experimental studies, and the preliminary
results coming from the NA49 experiment 
\cite{Roland} have found that the  fluctuations 
 (in the appropriate variable) follow
perfect Gaussian \footnote{ No non-statistical ``tails''
deviating from it was found, although Gaussian
was followed for several orders of magnitude.
} with small width,  clearly excluding 
any model based on superposition of independent
pp collisions.

This impressive
example have initiated the present work, which also 
 proposes several observables for which analysis of
 fluctuations is potentially useful for uncovering the details
of the system's evolution. However, our approach is more general,
it is not based on any particular model but rather on well known theory
of thermodynamical fluctuations. Although the predicted fluctuations are
probably not so spectacular as some dynamical models suggest, they are
guaranteed to be there. 

 2.In general, any statistical fluctuations can be derived from
the famous Boltzmann expression relating entropy S and probability P,
rewritten by Einstein  in the following form
\be P\sim exp(S) \ee
  Before we turn to specifics, few general comments about
it are appropriate. Applying general thermodynamical theory of fluctuations
 we rely on 
   statistical independence of fluctuations in different volume elements.
The fluctuations of temperature or
  particle composition should
be $independent$ on many dynamical details: it is irrelevant
whether all elements of the excited system have their freeze-out
at the same time or not, whether 
they freeze-out  at rest  in the same coordinate frame or
are all moving with different velocities\footnote{
 Still it is useful to imagine that the system
is cut
into many small (but still macroscopic) elements, and if each is taken
when it has a given $T,\mu$ and
put to rest and combined into a common system, 
the whole system will occupy some total volume V.}
The only thing which matters is $how$ the global entropy of the system S depends on the particular
observable under consideration.

  In general,
 the fluctuations  predicted by (1) neither
should   be   small, nor the
discussed system should have macroscopic number of degrees of freedom.
The only requirement is that the system is $equilibrated$, in the
sense that it equally
populates all its available phase space. 
Applications  of theory of thermodynamical fluctuations 
to  multi-hadron production reaction have a long history, going for pp
and $\bar p p$ reactions back to 70's. In 
\cite{Shu_72} 
probability of many  exclusive channels  for low energy
$\bar p p$ annihilation were calculated: simple textbook formulae
for the entropy 
have predicted quite accurately their probabilities,
 starting  with channels with only 4 secondaries\footnote{
Recall that the phase space for 4 particles is already a 8-dimensional
integral.
Using thermodynamical expressions instead of exact the phase space
means saddle point approximation: it 
has about the same accuracy as Sterling formula for the factorial $N\!\approx N
log(N/e)$,
which is excellent even for not-so-large N. 
}!
In  \cite{Shu_74} the production of $\bar K K $ pairs in
a pion gas was considered along these lines further. Summing
thermodynamical
formulae  over
states with only proper quantum numbers (no strangeness), a general
form of a
general dependence of the $K/\pi$ ratio on the system size
(the pion multiplicity $N_\pi$) was derived. It changes from 
 $O(N_\pi)$ value  for small enough systems to the $N_\pi$-independent
 ratio
for large ones, and describes  the data perfectly. Very recently similar
approach was used in ref.\cite{Becattini}, where excellent
 thermodynamical description of the yields of impressive list
of secondaries in  $e^+e^-,pp$ and nuclear collisions has been 
demonstrated\footnote{
We still do not understand why $e^+e^-,pp$ look so
equilibrated (see e.g. \cite{BH} for recent discussion)
. Heavy ion collisions is a different matter: they produce
larger and longer-lived systems, and also display other features of
 a macroscopic
behavior such as the transverse flow. }.

  3.Before discussing statistical fluctuations in details, let us make
  a general comment about their size.
For heavy ion reactions in question 
the multiplicity is $N= O(10^2-10^4)$.
 Therefore, the total entropy produced is  large and
the relative statistical
fluctuations are  small. Obviously they
scale   as $1/\sqrt{N}$ and are expected to be at few
percent level. Sufficient experimental statistics to measure such
fluctuations is not a problem.
 
In a standard way, the mean square deviations of
 various quantities can be obtained from  the expansion of the entropy
in powers of the deviation up to second order: thus proper
$succeptibilities$ 
appear. So, if  measured, the coefficients of $1/\sqrt{N}$   can 
  provide useful information about thermodynamics
 of hadronic matter.

    Our first example  is the  fluctuations in 
 temperature, which was recently discussed in ref.\cite{Stodolsky}\footnote{
I apologize that I did not know about it while
preparing the  preprint form of
this work: let me  thank  S. Mrowczynski who informed me.  
}.
 Standard thermodynamics
(see e.g. \cite{LL})  tells us that
\be P\sim exp[-{C_v(T)\over 2}(\Delta T/T)^2] \ee
%
where $\Delta T$ is the deviation of temperature from its mean value and
  $C_v(T)$  is the heat capacity of a hadronic matter.
(The heat capacity which enters here is extensive quantity $C_v=T{\partial S\over
\partial T}|_{N,V}$, proportional to the total volume V or the
number of particles N in the system, so the relative
 fluctuations  depend on  N as expected.) So in general, by measuring
 fluctuations in T we may learn about the heat capacity. 

   Let us now look at this general idea in more details, discussing
   which temperature we actually measure and what behavior of the
  fluctuations we may expect. If the temperature is deduced from
slopes of the observed $m_t=\sqrt{p^2+m^2}$ spectra, it corresponds to
the so called {\it thermal freeze-out} conditions\footnote{Not to be
  confused with those for {\it chemical freeze-out} ones at which
  particle
composition is determined. The former is related to elastic
scattering rates, the latter to inelastic species-changing ones. }.
It happens when the expansion rate of the system becomes comparable to
re-scattering rate. Note however, that for different secondaries the
re-scattering rate is rather different: say for SPS conditions it is
larger for nucleons than for pions, and is even smaller for kaons.
  Therefore one should see different 
temperatures, growing from $N$ to $\pi$ to $K$. This is indeed what
happens if one study  
 realistic cascades  (e.g.RQMD).

 Recently detailed study of thermal freeze-out have been also made
in hydro-based framework   \cite{HS_97}. In contrast to naive
freeze-out conditions used in most earlier hydro works
(freeze-out surfaces approximated by $isoterms$ $T=T_f$)
proper kinetic condition \cite{BGZ_78} (see also \cite{MS,HLR}) was
used, separately for $\pi,K,N$.
It was found that for different secondaries and collisions
 those surfaces have  quite different shapes. In particular,
the system cools deeper at its center, and remains hotter at edges: so
the relevant final $T_f$ is by no means a universal constant.
  Furthermore, they especially depend on the absolute
 system size: those which are produced with AA collisions with larger
 A cools
 to lower $T_f$. For example, 
for medium-size nuclei (Si-S)  these results suggest  $T_f=140-150 MeV$,
while for central heavy ion collisions (AuAu at AGS and PbPb at SPS)
 the  hadronic re-scattering goes on for long time (about 10-15 fm/c)
 and typical
$T_f$ becomes
 as low as 100-120 MeV.  
 Additional test of these numbers was provided
by a  detailed study of
the radial collective flow \cite{HS_97} and analysis of HBT radii
\cite{H_HBT} coming from NA49 experiment.

   Let us now return to fluctuations. Experimentally one should better
   look at some   simple observable, e.g. 
 deviations from the  mean $p_t$  \cite{GM_92}
or a fitted slope of the $m_t$ distribution $\tilde T$ in a event.
As it is well known, $\tilde T$ is not temperature but
  a combination of freeze-out local temperature $T_f$
and transverse flow velocity $v_t$ of matter elements, averaged over
the whole system.  It is complicated in practical calculations,
but we think the actual situation is significantly simplified because
  {\it 
the fluctuations of  $v_t$ are likely to be
 less important then those of  $T_f$}.
This    conjecture is motivated as follows:
(i) Unlike $T_f$ (which depends on random statistical
separation of total energy into transverse momenta, longitudinal momenta and
particle masses at the freeze-out moment),  $v_t$ is
basically an accumulated acceleration
during the whole evolution. Random fluctuations in pressure  should be
significantly washed out after the time integration.
 (ii) Furthermore, at AGS/SPS energies most
of the time the expanding system spends near the QCD phase transition,
in the so called mixed phase. As a result, most of the time available
 the pressure is nearly constant and  cannot
 significantly fluctuate.
(iii) 
The relative role of $T_f$ and $v_t$ in the observed slope
significantly
 depends on
the particle mass:  $v_t$ is much less important for pions than for
nucleons or deutrons. One may compare fluctuations of $\pi,N$
slopes (the main secondaries) to test our conjecture.
(iv)
  Finally, for pions corrections for transverse motion
is basically a ``blue shift'' factor, which cancels in ratios.

 This
conjecture
implies that 
\be 
({\Delta \tilde T\over  \tilde T})^2 \approx  ({\Delta  T_f\over
  \tilde T_f})^2={1 \over C_v(T_f)}
\ee
   The key observation at this point  is that because  $C_v(T)$
 has strong T-dependence, it can be used as
a ``thermometer" to test matter properties at thermal freeze-out.

  Particular form of the T-dependence of  the heat capacity
 in the vicinity of the QCD phase transition ($T\sim T_c\approx 150 MeV$)
depends on the order
of the phase transition\footnote{Delta function for the 1-st order, divergent
peak for  the second order and a finite peak
for a rapid cross-over.}. Large peak at $T=T_c$ has been observed
in lattice numerical data \cite{latticethermo}, but its exact
form\footnote{It depends on parameters like quark masses used and other methodical
details like
Wilson versus Kogut-Susskind fermions used.}
 and extrapolation to real QCD
is not yet clear.
   In the hadronic phase $T<T_c$ the
thermodynamics is usually described as a ``resonance gas"
\cite{resonancegas}.
As shown in \cite{HS_97}, the corresponding EOS agrees well with
 what is used in cascades like RQMD, and also explains the observed flow.
For practical purposes its T-dependence can be conveniently parameterized 
as $\epsilon, p \sim T^{1+1/c^2}$ where parameter $c^2$
(square of the sound velocity) is (nearly)
T-independent. It still depends on matter composition though
(baryon charge/entropy ratio), changing from  $c^2\approx .14$ for
baryon-rich matter produced at AGS  to $c^2\approx .2$ for baryon-poor one at SPS. 
Such EOS leads to  $C_v \sim T^{1/c^2}\sim T^{7}$ ($T^{5}$) at AGS
(SPS), respectively: so  even in the hadronic phase 
the T-dependence of $C_v$ is  strong  
enough to make a good thermometer.

  Let us now discuss how it is supposed to work.
At qualitative level, a debate continues on   whether the observed pions
are emitted  directly from the QGP clusters/mixed phase
 (see e.g. 
\cite{KC,PBM_etal}) at $T \approx T_c$, or (as we advocated above) they
cool further into a resonance gas phase (especially for heavy ions, Pb
or Au). In the former case one should find
 $C_v(T)$ at its peak:
the temperature fluctuations  should then be strongly suppressed.
In the latter scenario, $C_v(T)$ is much smaller, and we end up with 
a (counter-intuitive!) prediction that fluctuations for heavy ions
 should be relatively\footnote{Recall that we are discussing a
   coefficient of $1/\sqrt{N}$.}  {\it stronger}   compared to those
 for medium ions.

  Using the EOS of the resonance gas and
numbers for freeze-out temperatures $T_f$ for medium and heavy ion
collisions mentioned above
  (\cite{HS_97}) we can make more quantitative predictions. 
\be  {[N <(\Delta T_f/ T_f)^2>]^{heavy} \over [N <(\Delta T_f/ T_f)^2>]^{medium}}
 \sim { C_v(T_f^{medium}) \over C_v(T_f^{heavy})}  \sim
  [{T_f^{medium} \over T_f^{heavy}}]^{1/c^2} \ee
Although the
 values of $T_f$ mentioned above
are not that different, 
this ratio should change by about factor 2,
from medium to heavy ions. Such drastic change should be easily observable.
 
 4. Now we proceed to fluctuations of a more detailed observable, the 
occupation of 
particular momentum bins in the histograms.
In purely statistical system 
the
fluctuations in the particle number are generally described by
\be <{\Delta N^2}>=T{\partial N \over \partial \mu}|_{T,\mu} \ee
For example, for classical ideal gas we got trivial Poisson statistics. 
Deviations from it can be induced {\it without interaction}, just
   by quantum statistics. For the ideal 
Bose-Einstein (Fermi-Dirac) gas the expression
above leads to enhanced (suppressed) fluctuations:
\be \label{BEfluct} {<{\Delta n_k^2}>\over <n_k>}=(1 \pm  <n_k>) \ee
So, by measuring deviations from  Poisson statistics (1 in the r.h.s)
 one can learn what is the mean quantum degeneracy  factor $<n_k>$. 

 Bose-Einstein (Fermi-Dirac) correlations in two-body distributions
of pions
(nucleons)  originating from heavy ion collisions have been
studied
for a long time and is the basis of pion (nucleon) interferometry
providing information about space-time picture of the source.
 Correlation of 3 and more particles should be
 stronger
but present in so small corner of the phase 
space that those have not been used so far.

Fluctuations in the distributions on event-by-event basis we discuss
is just
  another way to approach the same ``induced radiation'' phenomenon.
The more bosons in a given momentum cell is produced in a given event,
the larger is the probability to find another one in the same cell.
Like for pion pairs, the interference happens late and
 the role of  interaction should be small.
    
  One of the problems with practical implementation of this simple
  idea is that
 the $total$ multiplicity of produced pions
is known to be a subject of large
non-statistical fluctuations,  in the first place
due to different impact parameters in
different events. There may be also other dynamical effects
\footnote{The multiplicity distribution 
in pp collisions are known to be wide and very non-statistical, and
there were even speculations that they
display fractal (intermittent) distributions.
 However, this is not so for nuclear collisions: the only
noticeable correlations observed so far
in this case are  due to Bose-Einstein correlations
 we discuss, see e.g\cite{Mike} .  }. Therefore one should
  consider deviations from
the normalized distributions, or $relative$ fluctuations in
  different
 bins.

  Pions are
 the only secondaries which are light enough to have
 noticeable quantum degeneracy  $n_k$ at freeze-out stage.
For chemically equilibrated pion gas
(chemical potential $\mu_\pi=0$) at zero momentum 
 Bose-enhancement factor $1+ n_k$ is increasing statistical
fluctuations 
(relative to random statistics) by a factor of about 1.5.
Naturally, the effect  disappears for large momenta. 

  Furthermore, it was  argued that  the
 pion gas is not chemically equilibrated at thermal freeze-out,
but has  a non-zero chemical potential\footnote{
The issue   has a somewhat
controversial history. It first surfaced because
of experimentally observed ``low $p_t$ enhancement''
relative to equilibrium ($\mu_\pi=0$) distributions.  
For discussion of 
pion ``over-population'' of   the phase space
see e.g. \cite{Pratt} and references therein: even a  ``pion laser'' conditions
( $ n_{k=0} \rightarrow \infty$) were discussed. 
However, in fact most of this enhancement
is  due to resonance decays.
} 
 $\mu_\pi$, so that
 the effect should be even stronger. The basic reason for
non-zero ($\mu_\pi$ 
was originally pointed out by G.Baym: adiabatic
expansion of the re-scattering pion gas creates it, even if one starts
from
 chemical
 equilibrium ($\mu_\pi=0$).
Estimates made in \cite{Bebie_etal,HS_97} has resulted in small
effect for medium ions, but for heavy ones (PbPb collisions at SPS)
it should  reach for mid-rapidities
 $\mu_\pi\sim 60 MeV$. Moreover, in \cite{HS_97} it is shown that
comparison of NA44 data for PbPb and
 SS shows this effect, with $\mu_\pi\sim 60 MeV$.
  If so,  
Bose enhancement 
 of fluctuations at small $p_t$ should be larger, giving the  $1+ n_k$
 factor of 2.0-2.3. 
Again this prediction looks
counter-intuitive: (relative) fluctuations are 
enhanced  for $larger$
(not smaller) systems!
 Finally, let us point out
direct  relation between this effect and one of the possible manifestation
of DCC. As discussed in \cite{DCC}, the remnants of the
disoriented vacuum can be vied as
 the unusually large population of the low-$p_t$ bins in the pion $p_t$
distribution. If this indeed happens, expression (\ref{BEfluct}) tells us that
one should also observe enhanced fluctuations in the corresponding bins induced
by Bose-Einstein interference effects.

  So far we have discussed an idealized picture of a gas of secondary
  pions:  now we are going to face real situation. First of all, 
discussion above implied that
 there exist some common reference frame in which it is at rest,
 so that 
 the ``small momentum'' bin discussed above
in unambiguous. In reality the pion system consists
of volume elements moving with different collective velocities,
and this restrict/dilute the 
 Bose 
enhancement effects. The density leading to enhancement is actually
density in the {\it phase space}:  even for pions with identical momenta this
effect is absent if their emission points  are too far apart in space
or
too much separated in time. 
This issue was much debated for two-pion interferometry, with
realization that the the so called
HBT radii $R_{x,y,z}$ are not the size of the system, but a size of a
region in which the velocity change (due to its gradient) can be
compensated
by thermal
velocities. (In other words, those are related to curvatures of the
freeze-out surface.)  Bose 
enhancement of fluctuations is as strong as discussed above, provided
 the bin used has the
size such that $|\Delta  p_i| R_i < 1 $. If for statistics reason
the bin is larger, the effect is reduced accordingly.

  The second problem is again well understood for two-pion
  interferometry.
Substantial fraction $f_{LL}$ 
 of secondary pions come from long-lived (LL) resonances like $\omega$ and
weak $K_0$ decays: those do not interfere with others (unless the
resolution
or bin size is as small as the inverse lifetime). Two-pion
interferometry and the Bose enhancement term we speak
about is reduced by  it by the same factor $(1-f_{LL})^2$.

 In summary, we have pointed out that  statistical fluctuations
of the collective observables are related to thermodynamically 
interesting quantities. In particular, the r.m.s. deviation in the apparent
temperature of the events is related to the heat capacity at freeze-out,
and fluctuations in the $p_t$ spectrum at small $p_t$ are 
sensitive to freeze-out temperature and pion chemical potential.  

Acknowledgments. This paper was triggered by the talk by G.Roland at
Hirschegg-97.  I also thank
M.Gazdzicki and S. Mrowczynski
for related discussion, and referees of the paper for their thoughtful
reading of the original manuscript.
This work  is partially supported by US DOE. 
\par

\end{document}